\documentclass[usenatbib,a4paper]{mn2e}
\usepackage{hyperref}
\usepackage{hypernat}
\usepackage{graphicx}
\usepackage{verbatim}
\usepackage{color}

\newcommand\kms{km$\,$s$^{-1}$}
\newcommand\Msol{M$_{\odot}$}
\newcommand\nfa{$N_\mathrm{FA}$}
\newcommand\mhi{$M_\mathrm{HI}$}

\begin{document}

\title[HIMF \& Environment]{Environmental dependence of the HI mass function in the ALFALFA 70\% catalogue}
\author[Jones et al.]{Michael G. Jones$^{1}$\thanks{E-mail: jonesmg@astro.cornell.edu}, Emmanouil Papastergis$^{2}$, Martha P. Haynes$^{1}$ \newauthor and Riccardo Giovanelli$^{1}$
\\
$^{1}$Cornell Center for Astrophysics and Planetary Science, Space Sciences Building, Cornell University, Ithaca, NY 14853, USA
\\
$^{2}$Kapteyn Astronomical Institute, University of Groningen, Landleven 12, Groningen NL-9747AD, Netherlands}

\maketitle

\begin{abstract}
We search for environmental dependence of the HI mass function in the ALFALFA 70\% catalogue. The catalogue is split into quartiles of environment density based on the projected neighbour density of neighbours found in both SDSS and 2MRS volume limited reference catalogues. We find the Schechter function `knee' mass to be dependent on environment, with the value of $\log ({M_{*}/\mathrm{M_{\odot}}})$ shifting from $9.81 \pm 0.02$ to $10.00 \pm 0.03$ between the lowest and highest density quartiles. However, this dependence was only observed when defining environment based on the SDSS reference catalogue, not 2MRS. We interpret these results as meaning that the local environment is the dominant cause of the shift in $M_{*}$, and that the larger scales that 2MRS probes (compared to SDSS) are almost irrelevant. In addition, we also use a fixed aperture method to probe environment, and find tentative evidence that HI-deficiency depresses the value of $M_{*}$ in the highest density regions. We find no significant dependence of the low-mass slope on environment in any test, using either method. Tensions between these results and those from the literature, are discussed and alternative explanations are explored.
\end{abstract}

\begin{keywords}
galaxies: mass function
\end{keywords}

\section{Introduction}

The HI mass function (HIMF) is the density distribution of HI masses of galaxies in the Universe and represents a key component in understanding how collapsed structures form. HI surveys are complementary to optical surveys, and the galaxy luminosity functions they deliver, because they have fundamentally different selection effects and thus detect a different component of the underlying galaxy population. Together the luminosity functions and mass functions that these surveys calculate offer important constraints on the population of galaxies that simulations of structure formation generate. 

Detailed studies of the HIMF have only become possible in the last decade or so, as previously sample sizes were too small and selection effects too poorly understood. With the advent of wide area, blind surveys like HIPASS \citep[HI Parkes All Sky Survey;][]{Barnes+2001} and ALFALFA \citep[Arecibo Legacy Fast ALFA survey;][]{Giovanelli+2005} precise determination of the HIMF in the local Universe has become possible, with both HIPASS and ALFALFA \citep{Zwaan+2005,Martin+2010} indicating that the HIMF is well fit by a Schechter function \citep[an analytic expression for the mass distribution of collapsed objects in an expanding universe,][]{Press+Schechter1974,Schechter1976}, with a low-mass slope of approximately -1.3 and a `knee' mass of almost $10^{10}$ \Msol. The large area and source counts of these surveys have also allowed studies of environmental dependence that are not restricted to 10s or 100s of objects and a handful of nearby groups.

Although many studies looking for environmental dependence have been carried out \citep[for example][]{Rosenberg+2002,Springob+2005,Zwaan+2005,Stierwalt+2009,Moorman+2014}, it is still important to ask why any environmental dependence is expected at all? There are many processes and properties that are known to depend on a galaxy's environment, here we will briefly discuss a few that we expect to be the most influential on a galaxy's HI content. First, due to their mass and tendency to cluster, more massive dark matter (DM) halos are generally found in more overdense regions. Thus, the `knee' mass ($M_{*}$) of a Schechter function fit to the HIMF, would be expected to increase towards more dense regions of the Universe. Secondly, voids can be considered as more slowly evolving sections of our Universe \citep{Peebles2001,Tinker+Conroy2009}. This means that by isolating the void galaxies in a sample, you are effectively probing the HIMF at a previous time, where systems are likely to be lower mass and more numerous, assuming a hierarchical model of galaxy formation. Therefore, it would be expected that the low-mass slope would steepen within lower density regions. In addition to these two effects, in the most dense regions (galaxy clusters) galaxies will be unable to retain their neutral gas due to the harassment and ram pressure stripping they experience, and so might be expected to be HI-deficient with respect to galaxies in the field; while galaxies in voids are likely more prone to background UV heating than those in the field \citep{Hoeft+2006}. Given all of the above, some environmental dependence in the shape of the HIMF is expected, however there are numerous competing affects, making the exact nature of the dependence difficult to predict. To complicate matters further, most studies have thus far produced marginal and/or conflicting results.

Using the Arecibo Dual Beam Survey \citep[ADBS;][]{Rosenberg+2000} \citet{Rosenberg+2002} found that the HIMF low-mass slope ($\alpha$) was flatter in Virgo than the $\sim$-1.5 value found in the rest of the survey. However, the paper points out that small number statistics and distance errors make their results somewhat uncertain. \citet{Springob+2005} also found (at low significance) that both $\alpha$ and $M_{*}$ decrease in high density environments, from their analysis of an optically selected sample from the Arecibo General Catalog \citep{Springob+2005b}. However, more recently, \citet{Stierwalt+2009} used an early ALFALFA release to show essentially the opposite result, that the low-mass slope in the dense Leo region was steeper than other measurements of the HIMF at the time \citep[though, given the quoted error, is now consistent with that of the global ALFALFA HIMF;][]{Martin+2010}. There are also a number or other results from surveys of individual groups \citep{Verheijen+2001,Kovac+2005,Freeland+2009,Pisano+2011} which generally imply that the low-mass slope is flatter in galaxy groups.

\citet{Zwaan+2005} concluded that $\alpha$ steepened in high density environments, based on data from HIPASS. However, unlike all other studies, the proximity to other HI galaxies was used to define environment (rather than an optically selected reference catalogue). HI surveys are known to be incomplete for galaxies in the densest environments, which combined with the fact that HIPASS is not a volume limited catalogue, makes a comparison with this result difficult; but we note that attempting to perform a similar experiment with ALFALFA did not result in any apparent environmental dependence in the HIMF. Most recently \citet{Moorman+2014} used the 40\% ALFALFA catalogue ($\alpha$.40) to search for environmental dependence based on void and wall regions defined using the method devised by \citet{Hoyle+Vogeley2002}. They found no evidence of any change in $\alpha$, but contrary to \citet{Springob+2005} $M_{*}$ was found to increase in denser regions. This represents the most statistically significant result of large scale environmental dependence in the HIMF to date, which is in part due to the greatly larger sample size that ALFALFA provides. Since that study, data from 30\% more of ALFALFA's nominal area ($\sim$7,000 deg$^{2}$) have been reduced, and $\sim$7,000 additional high signal-to-noise HI sources have been extracted.

In this paper we choose to focus on a local definition of galaxy environment, rather than defining voids, walls and clusters, for two reasons. First, because the majority of the additional 30\% added to the ALFALFA catalogue since the \citet{Moorman+2014} study is not within the SDSS (Sloan Digital Sky Survey) spectroscopic footprint, making defining voids problematic; and secondly because related optical and theoretical works \citep{Berlind+2005,Blanton+2006,Tinker+Conroy2009} find that galaxy properties are most closely related to a galaxy's host halo, and may even be almost independent of its large scale environment. Obviously the two are not independent, but if a galaxy's properties depend mostly on its host halo mass rather than its ``assembly bias" \citep[the idea that haloes of a given mass, but which assemble at different times, will cluster differently, e.g. see][]{Wechsler+2006}, then the strongest signal of any change in the mass function would presumably arise from a measure of local environment, rather than large scale structure (LSS).

We use a combination of SDSS data release 8 \citep{Aihara+2011} and the 2MASS Redshift Survey \citep[2MRS;][]{Huchra+2012} as reference catalogues to define the local density of ALFALFA galaxies based on the separation of their projected nearest neighbours in these catalogues. This allows us to split the HI sources into quartiles of differing environment and calculate the HIMF for each environment separately. 2MRS allows us to make use of the full ALFALFA 70\% sample, while the superior depth of SDSS permits smaller scale environments to be probed.

In the following section we give a brief overview of the ALFALFA survey, in \S\ref{sec:env} we describe our definitions of environment, \S\ref{sec:HIMFcalc} outlines how the HIMF is calculated, and our results are presented in \S\ref{sec:results}. The implications of these results are discussed in \S\ref{sec:discuss}, and finally we draw our conclusions in \S\ref{sec:conclude}.

\section{The ALFALFA sample}
\label{sec:alfalfa_sample}

Observations for the main ALFALFA survey were completed in October 2012 after over 7 years of observing with the 305 m Arecibo radio telescope in Puerto Rico. The final ALFALFA footprint covers approximately 6,900 deg$^{2}$ on the sky, and is broken up into two contiguous regions: one ranges from $\sim$7.5 hr RA to $\sim$16.5 hr RA in the Arecibo Spring sky, and the other from $\sim$22 hr RA to $\sim$3 hr RA in the Arecibo Fall sky. While the Spring ALFALFA region has almost complete overlap with SDSS spectroscopy, in the Fall sky there are only a few stripes where spectra are available. The drift scan observing strategy of ALFALFA proved extremely successful with over 95\% of observing time spent with the ``shutter open", including all start-up, shutdown and calibration procedures. A matched filtering algorithm \citep{Saintonge+2007} is used to help identify sources, but all ALFALFA spectra are ultimately extracted by a person, and the current progress is over 70\% complete, yielding over 20,000 high signal-to-noise (S/N) sources and counting. Over 99\% of these HI sources have identified optical counterparts (with matching redshifts where optical spectra exist).\footnotemark{}

\footnotetext{The ALFALFA 70\% catalogue is publicly available at \url{http://egg.astro.cornell.edu/alfalfa/data/index.php}}

In order to calculate the HIMF it is essential to have HI masses for the ALFALFA sources, which in turn necessitates distance measurements for every source. ALFALFA uses a flow model developed by \cite{Masters2005} to convert recessional velocities below 6,000 \kms \ to distances. Distances to galaxies beyond 6,000 \kms \ are calculated instead assuming Hubble flow, with $H_{0}$ = 70 \kms$\,\mathrm{Mpc^{-1}}$. In addition, 303 sources are assigned to regions of the Virgo cluster by matching to the VCC \citep[Virgo Cluster Catalog;][]{Binggeli+1985}, 1,130 sources are assigned to groups from 2MRS \citep{Crook+2007} and given the mean velocity of the group members, and 63 (1,646) sources are given their primary (secondary) distances from the literature. Note that in this article we only consider galaxies in the 70\% ALFALFA catalogue within the range of distances 1,000-15,000~\kms$/H_0$.

Once distances to ALFALFA galaxies have been calculated, their HI masses can be computed through the usual equation:
\begin{equation}
\frac{M_{\mathrm{HI}}}{\mathrm{M}_{\odot}} = 2.356 \times 10^{5} D_{\mathrm{Mpc}}^{2} S_{21} \;\; .
\end{equation}
In the equation above, $D_{\mathrm{Mpc}}$ is the distance to the galaxy in Mpc and $S_{21}$ is its integrated flux in Jy \kms.

\section{Quantifying Environment}
\label{sec:env}

The term `environment' has no objective definition, and different studies have used drastically different methods to describe it quantitatively. On one extreme we can find techniques that characterise the environment based on the morphology of the cosmic web, classifying galaxies as void, wall, and filament objects \citep[e.g.][]{Hoyle+Vogeley2004,Rojas+2004,Hoyle+2005}. On the other extreme, it is possible to characterise the most immediate surroundings of a galaxy based on its status as a central or satellite galaxy \citep[e.g.][]{Carollo+2013}. In this article we choose to study the dependence of the HIMF on the \textit{local} environment of ALFALFA sources, as traced by the proximity of neighbouring galaxies. More specifically, we employ the widely-used nearest neighbour (NN) and fixed aperture (FA) methods to quantify the environment \citep[e.g.][]{Muldrew+2012}. The former method calculates a local density based on the distance between the target galaxy and its $N^\mathrm{th}$ nearest neighbour. The latter is instead based on the number of objects found within a region of fixed size surrounding the target galaxy.

Each method of environment characterisation has its own set of advantages and drawbacks, and there are often trade-offs between a method's physical motivation and its simplicity. Our choice to use the NN and FA methods is based on the fact that these two methods are purely observational, and have a clear and intuitive definition. Sections \ref{sec:reference_catalogue}--\ref{sec:fixed_aperture} below contain a detailed description of the methods' implementation in the context of the ALFALFA sample.

\subsection{An external reference catalogue for environment characterisation}
\label{sec:reference_catalogue}

\begin{figure*}
\centering
\includegraphics[width=\textwidth]{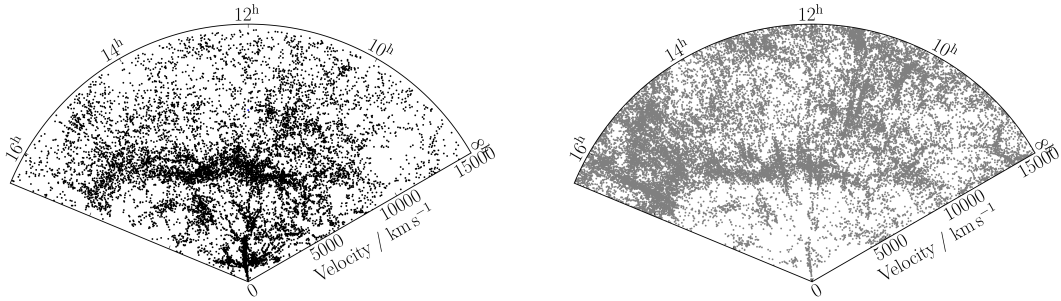}
\caption{
\textit{Left panel}: Coneplot of ALFALFA galaxies in the Spring region of the sky. \textit{Right panel}: Coneplot of the SDSS galaxies in the reference volume-limited catalogue, within the same volume as the ALFALFA sample. The environment of each ALFALFA galaxy in the left panel is calculated based on the position of neighbours in the reference catalogue shown in the right panel (refer to \S\ref{sec:reference_catalogue} for details).  
}
\label{fig:a70+SDSS_cone}
\end{figure*}
The simplest way to find neighbouring galaxies for the ALFALFA sources would be to search within the ALFALFA catalogue itself. This approach has been previously used by \citet{Zwaan+2005} to measure the environment of galaxies detected by the HIPASS blind HI survey. Even though straightforward, this methodology comes with two important observational disadvantages. First, any blind HI survey produces a nearly flux-limited\footnotemark{} sample. As the left panel of Figure \ref{fig:a70+SDSS_cone} shows, the number of detections in such a sample drops in the outer parts of the survey, since only the most HI massive galaxies remain visible at these large distances. Consequently, a bias is introduced in the measurement of environment, whereby galaxies appear systematically more isolated with increasing distance. Second, galaxies located in the central regions of clusters and rich groups are known to be HI-deficient with respect to their peers in the field \citep[][for a review]{Haynes+1984}. This means that HI-selected samples are biased against the highest density regions of the cosmic web. This effect can be clearly seen either directly in the spatial distribution of ALFALFA galaxies near clusters (see figure 6 in \citealp{Haynes+2011}), or indirectly in the clustering properties and the colour-magnitude diagram of ALFALFA galaxies (see figure 20 in \citealp{Papastergis+2013} and figure 10 in \citealp{Huang+2012b}, respectively).

\footnotetext{In reality, the detection limit of a blind HI survey depends both on the integrated flux and the width of a galaxy's HI profile (see section 6 in \citealp{Haynes+2011}). However, the width dependence of the detection limit is mild enough such that the detectability of a galaxy by ALFALFA depends primarily on its HI mass.}

In this article we remedy these shortcomings by defining the environment of ALFALFA galaxies based on an external reference catalogue. The catalogue we use has two important properties:

\begin{enumerate}

\item \textit{It is optically selected.} In particular, we use galaxies from the spectroscopic database of the eighth data release of the Sloan Digital Sky Survey (SDSS DR8; \citealp{Aihara+2011}). This property ensures that we trace the environment well even in high density regions where gas-deficiency becomes an issue.     

\item \textit{It is volume-limited.} We include in the reference catalogue only galaxies that are brighter than $M_r = -18.9$. Given the apparent magnitude limit for the SDSS spectroscopic sample ($m_r = 17.75$) and the maximum distance cut for the ALFALFA sample ($\approx$214 Mpc), these galaxies are bright enough to constitute a volume-complete sample within the ALFALFA volume. In turn, this ensures that environment is measured consistently regardless of the distance at which the ALFALFA galaxy is located.

\end{enumerate}

\begin{figure}
\centering
\includegraphics[width=\columnwidth]{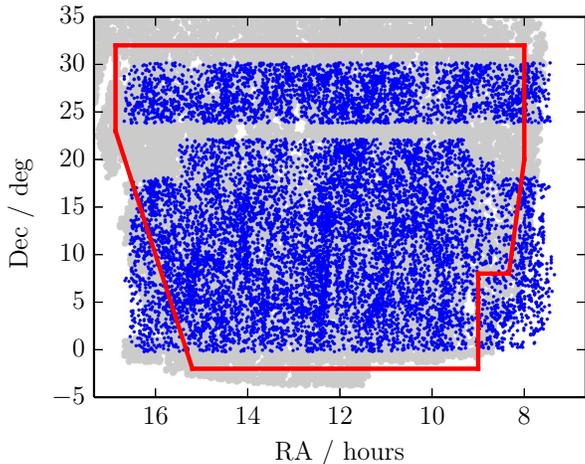}
\caption{The sky positions of the sources in the ALFALFA 1,000-15,000 \kms \ sample (small blue points), and the 500-15,500 \kms \ SDSS reference catalogue (large, overlapping grey points). The thick red line is the cut that is applied to the ALFALFA sample when comparing with SDSS, in order to ensure there is more than complete overlap.}
\label{fig:sky_cov}
\end{figure}

The right panel of figure \ref{fig:a70+SDSS_cone} shows the spatial distribution of the SDSS reference catalogue. As expected from its volume-limited nature, the number of objects in the reference catalogue grows steadily with increasing distance. Note that in order to avoid edge effects, the reference catalogue is slightly more extended in the radial direction than the ALFALFA sample, covering the distance range 500 -- 15,500~\kms$/H_0$. We remind the reader that distance cuts are quoted in terms of recessional velocity, but they actually refer to distances that are estimated as described in \S\ref{sec:alfalfa_sample}. In order to avoid edge effects in the plane of the sky as well, the reference catalogue must have more than complete sky overlap with the ALFALFA sample. Figure \ref{fig:sky_cov} shows the footprints of the ALFALFA sample and the SDSS reference catalogue in the Spring region of the sky, and details the complicated sky mask that is necessary to maximise the number of ALFALFA galaxies while maintaining high levels of overlap with the reference catalogue. Keep in mind that, given the poor spectral coverage of SDSS in the Fall region of the sky, a different reference catalogue is necessary to study the 70\% ALFALFA sample over its full sky extent (see \S\ref{sec:2MRS}). 

\begin{figure}
\centering
\includegraphics[width=\columnwidth]{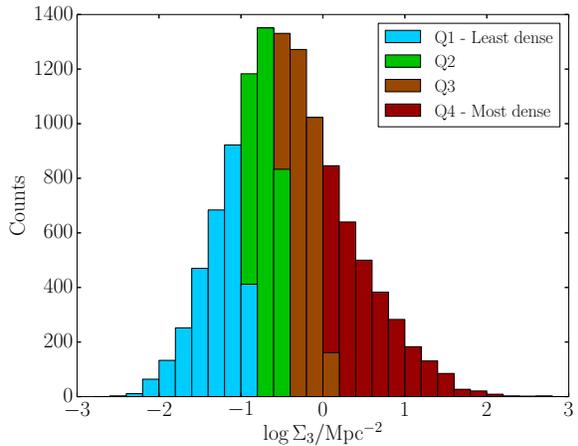}
\caption{
Histogram of the 3rd nearest neighbour density, $\Sigma_3$, for ALFALFA galaxies. The density of each ALFALFA galaxy is calculated based on the proximity of neighbouring objects in an SDSS volume-limited reference catalogue (refer to \S\ref{sec:reference_catalogue} \& \S\ref{sec:nearest_neighbour}). Different colours and hatching styles mark the four quartiles of the distribution, which from light blue to dark red (light to dark colours, and left to right) contain galaxies situated in progressively denser environments.
}
\label{fig:NN3_hist}
\end{figure}

Defining environment in this way, based on a volume-limited reference catalogue avoids the need to place harsh flux cuts on the ALFALFA sample (to make it volume-limited), as its sensitivity and completeness are well understood \citep{Haynes+2011} and can be corrected for independently of our external definition of environment, as will be described in \S\ref{sec:HIMFcalc}.

\subsection{Nearest Neighbour Environment}
\label{sec:nearest_neighbour}

\begin{figure*}
\centering
\includegraphics[width=\textwidth]{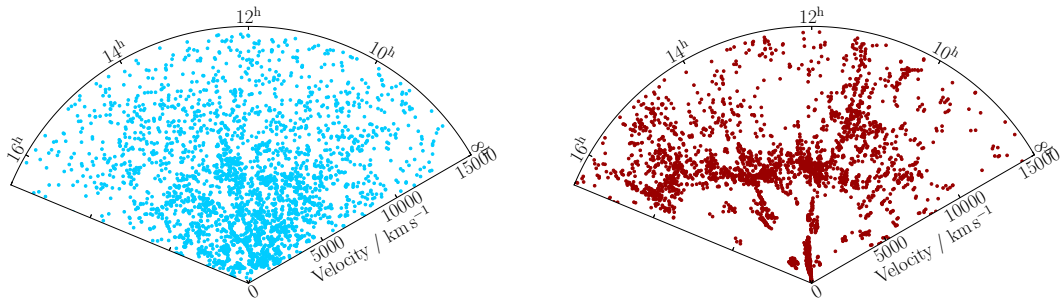}
\caption{
Coneplots of ALFALFA galaxies belonging to the lowest density quartile (\textit{left panel}) and highest density quartile (\textit{right panel}) of the nearest neighbour density distribution (see figure \ref{fig:NN3_hist}). Note the marked difference in clustering between these two environmental subsamples.
}
\label{fig:SDSS_NN_cone}
\end{figure*}

We calculate a nearest neighbour density for each ALFALFA galaxy based on the projected distance to the third closest galaxy in the reference SDSS catalogue. First, we record the sky position of all objects in the reference catalogue that have a recessional velocity within $\pm$500 \kms \ from the recessional velocity of the target ALFALFA galaxy. We then identify the third nearest object in the plane of the sky, and calculate its projected separation at the distance of the ALFALFA galaxy, $R_3$. The projected nearest neighbour density can then be calculated as    

\begin{equation}
\Sigma_{3} = \frac{3}{\mathrm{\pi} R_{3}^{2}} \;\;\; . 
\end{equation}

When identifying neighbours, we exclude any object in the reference catalogue that is located within 5 arcsec and $\pm$70 \kms \ from the ALFALFA galaxy; such an object corresponds (almost always) to the counterpart of the ALFALFA galaxy in SDSS. Throughout this article, $\Sigma_{3}$ will be used to characterise the local environment via the NN method, and will often be referred to as simply `the environment' or `local density'.

Figure \ref{fig:NN3_hist} shows the distribution of $\Sigma_3$ for the ALFALFA galaxies. Based on the distribution's approximately lognormal shape, we divide the ALFALFA sample into four quartiles which contain objects residing in increasingly denser environments. Figure \ref{fig:SDSS_NN_cone} shows coneplots of the ALFALFA galaxies belonging to the lowest and highest density quartile (left and right panel, respectively). Reassuringly, the difference in clustering between the two environmental subsamples is clearly visible by eye. Sources in the densest environment are grouped together in clumps and filaments, whereas the sources in the least dense environment are distributed almost uniformly in space. This is an excellent indication that the NN method is splitting the ALFALFA galaxies into environmental subsamples in a sensible way.

\subsection{Fixed Aperture Environment}
\label{sec:fixed_aperture}

\begin{figure}
\centering
\includegraphics[width=\columnwidth]{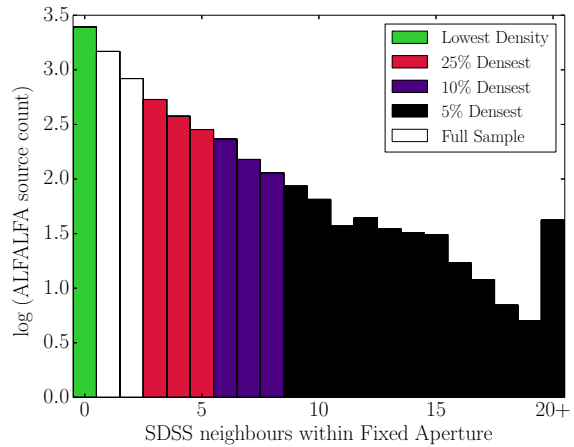}
\caption{
Histogram of the number of SDSS neighbours within the fixed aperture, \nfa, for galaxies in the ALFALFA sample (see \S\ref{sec:fixed_aperture}). The green (leftmost) bar denotes the lowest density subsample, \nfa$=0$. The crimson, purple and black bars (left to right) represent instead the ALFALFA galaxies located in the densest 25\%, 10\% and 5\% environments, according to the fixed aperture method (\nfa$\geq3$, \nfa$\geq 6$ and \nfa$\geq 9$, respectively). Note that these three all overlap as the densest 25\% includes both the densest 10\% and 5\%. The final bin contains counts for all ALFALFA sources with 20 or more SDSS neighbours within the fixed aperture. The white bars correspond to galaxies with $0<$\nfa$<3$.
}
\label{fig:FA_hist}
\end{figure}

In addition to the NN method described above, we also adopt a fixed aperture approach as a complementary way to measure the environment of ALFALFA galaxies. In particular, we count the number of galaxies in the reference catalogue that lie within a radius of 1 Mpc and a velocity range of $\pm$500 \kms \ from the position and velocity of our target ALFALFA galaxy. The fixed aperture environment is thus characterised simply by a natural number, $N_\mathrm{FA}$. As with the nearest neighbour method, we exclude possible optical counterparts from the count (any object that is within 5\arcsec \ and $\pm$70 \kms \ from the ALFALFA galaxy).

\begin{figure*}
\centering
\includegraphics[width=\textwidth]{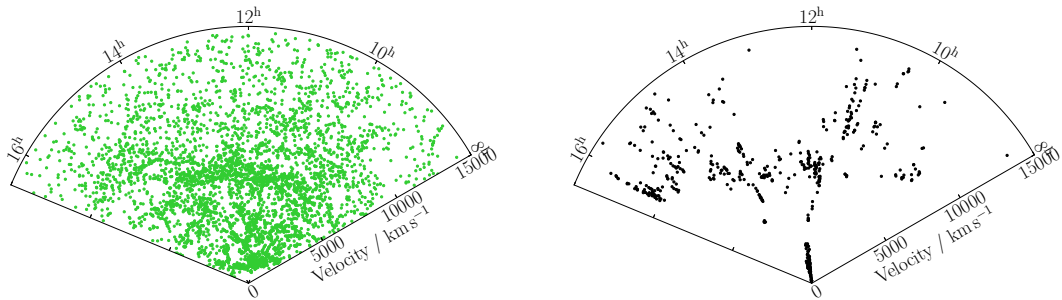}
\caption{
Coneplots of ALFALFA galaxies belonging to the lowest density subsample (\textit{left panel}) and highest density subsample (\textit{right panel}) of fixed aperture environment (refer to figure \ref{fig:FA_hist}). The latter sample demonstrates that the fixed aperture method can be used to probe the largest groups and clusters in the survey volume.
}
\label{fig:FA_cone}
\end{figure*}

Figure \ref{fig:FA_hist} shows the distribution of fixed aperture environment, \nfa, for the ALFALFA sample. Unlike in the case of nearest neighbour densities, the distribution of \nfa \ has a power law form. This means that the fixed aperture method provides a rather coarse description of environment at low densities; for example, the lowest FA density subsample (\nfa $=0$) contains already 38\% of the total sample. On the other hand, the FA method is better for isolating the ALFALFA galaxies that reside in the highest density environments. The preceding points are visually demonstrated by the two coneplots in Figure \ref{fig:FA_cone}: The left panel shows the lowest FA density subsample of ALFALFA. This sample is successful at tracing low density environments in general, but cannot discriminate between galaxies located in voids and galaxies located in parts of filaments with low local density. On the other hand, the right panel shows the top 5\% of ALFALFA galaxies in terms of FA density (\nfa $\geq 9$). This latter sample does an excellent job at tracing the locations of the largest clusters and groups in the survey volume.

\subsection{2MRS Nearest Neighbour Environment}
\label{sec:2MRS}

In order to study the environment of the 70\% ALFALFA sample in both the Spring and Fall regions of the sky, we need a reference catalogue that covers the entire celestial sphere. To this end, we follow the same approach described in \S\ref{sec:reference_catalogue}, but now using the all-sky 2MASS Redshift Survey (2MRS) as reference. We select galaxies in the 2MRS that are brighter in the $K$-band than $M_K = -24.9$. Given the 2MRS apparent magnitude limit of $m_K = 11.75$, this cut makes the 2MRS catalogue volume-limited over the entire volume probed by ALFALFA out to 15,000 \kms/$H_{0}$. 

Compared to the SDSS spectroscopic survey, the 2MRS survey is much shallower. This means that the 2MRS-based reference catalogue is limited to much brighter objects than the SDSS-based one, and consequently it is much sparser in space. This fact affects the way in which environment is measured with the nearest neighbour method. In particular, the third nearest neighbour in the 2MRS catalogue is usually so far apart from the target ALFALFA galaxy that it does not provide a good measure of local environment. As a result, when using the 2MRS catalogue as reference we calculate local densities based on the distance to the nearest neighbour, $R_1$; the corresponding density is then $\Sigma_1 = 1/\pi R_1^2$.

\begin{figure}
\centering
\includegraphics[width=\columnwidth]{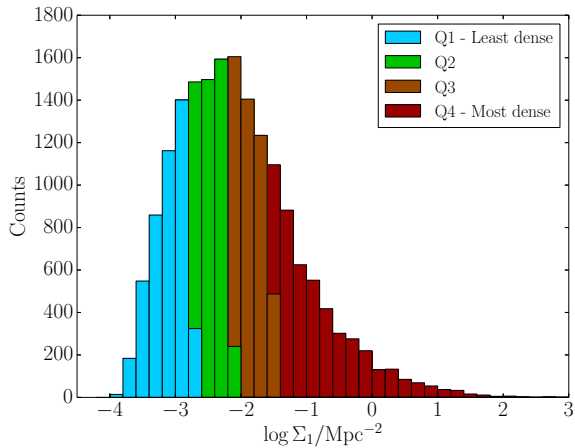}
\caption{
Similar to figure \ref{fig:NN3_hist}, but referring to the environment as defined by the 2MRS reference catalogue. Keep in mind that in the case of 2MRS the nearest neighbour density is calculated based on the distance to the closest neighbour (\S\ref{sec:2MRS}). Once again, different colours (shades) mark the four quartiles of the distribution, increasing in density left to right.
}
\label{fig:NN1_hist}
\end{figure}

\begin{figure*}
\centering
\includegraphics[width=\textwidth]{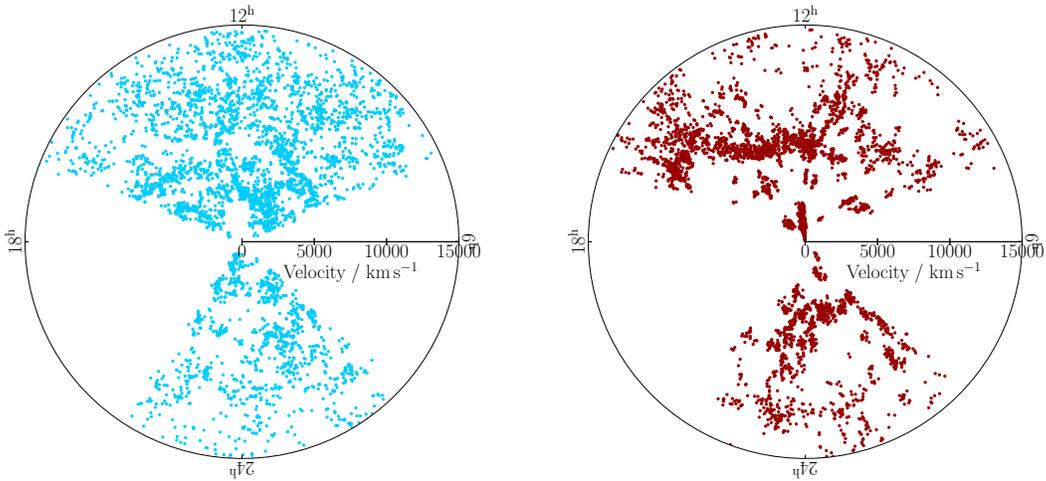}
\caption{
Same as figure \ref{fig:SDSS_NN_cone}, but showing coneplots of the lowest (\textit{left panel}) and highest (\textit{right panel}) density quartiles of the 2MRS nearest neighbour environment (see figure \ref{fig:NN1_hist}). Note that the 2MRS is an all-sky survey, and therefore it can be used to measure the environment in both the Spring and Fall portions of the ALFALFA footprint.
}
\label{fig:2MRS_NN_cone}
\end{figure*}

Figure \ref{fig:NN1_hist} shows the distribution of $\Sigma_1$ for the full ALFALFA sample. We follow the same process described in \S\ref{sec:nearest_neighbour} and split the distribution into four quartiles, containing galaxies located in progressively denser environments. Note that despite the change in nearest neighbour rank, the scale over which environment is probed by 2MRS is still larger than in the SDSS case. This is evident by the shift in the location of the distribution peak between figures \ref{fig:NN3_hist} and \ref{fig:NN1_hist}; the former peaks at $\sim$0.16 Mpc$^{-2}$, and the latter at $\sim$0.01 Mpc$^{-2}$. The difference in scale over which environment is probed is also reflected in the spatial distribution of the 2MRS environmental subsamples. This is clearly visible in Figure \ref{fig:2MRS_NN_cone}, which plots the spatial distribution of the lowest and highest density 2MRS quartiles (left and right panel, respectively). By comparing with the corresponding panels in figure \ref{fig:SDSS_NN_cone}, one can immediately recognise that the 2MRS environmental subsamples follow more closely the cosmic LSS than their SDSS counterparts. For example, large filaments are more starkly defined in the highest density 2MRS sample than in the highest density SDSS sample. At the same time, galaxies in the lowest density 2MRS sample actively avoid the locations of large filaments, an effect that is not present in the corresponding SDSS sample (see figure \ref{fig:SDSS_NN_cone}).

\section{Calculating HIMFs}
\label{sec:HIMFcalc}

The HI mass function (HIMF) is defined as the number density of galaxies as a function of their HI mass, $\phi(M_{HI})$. Galaxies span several orders of magnitude in terms of their HI mass, so the HIMF is customarily measured in logarithmic mass intervals as 

\begin{eqnarray}
\phi(M_\mathrm{HI}) = \frac{dN_{\mathrm{gal}}}{dV \: d\log_{10}(M_\mathrm{HI})} \;\;\; .
\end{eqnarray}

\noindent
In the equation above, $dN_{\mathrm{gal}}$ is the average number of galaxies in a cosmic box of volume $dV$, whose HI mass lies within a small logarithmic bin centred around $M_{\mathrm{HI}}$.

Since the ALFALFA sample is (roughly) flux-limited, the measurement of the HIMF is not a simple counting exercise. For example, there are many more detections in ALFALFA with \mhi $= 10^{10}$ \Msol \ than with \mhi $= 10^8$ \Msol, but the former sources can be detected out to much larger distances than the latter. Once the sensitivity limits of the survey are known \citep[section 6]{Haynes+2011}, this effect can be compensated for by weighting each source according to the maximum volume over which it is detectable by the survey (`$1/V_\mathrm{max}$' method).

The $1/V_\mathrm{max}$ method has the advantage of being intuitive and simple to implement, but has one major limitation: it is unbiased only if the galactic population is distributed in an approximately uniform way within the survey volume. This is definitely not the case for the ALFALFA survey, where large-scale structure is clearly present in the spatial distribution of galaxies (see figure \ref{fig:a70+SDSS_cone}). For this reason, we use in this article a more sophisticated method to calculate the HIMF, referred to as the `$1/V_\mathrm{eff}$' method \citep{Zwaan+2005}. More specifically, the HIMF can be calculated within logarithmic mass bins as

\begin{eqnarray}
\phi_i = \frac{1}{\Delta m_\mathrm{HI}} \cdot \sum_j \frac{1}{V_{\mathrm{eff},j}} \;\;\; , 
\end{eqnarray}

\noindent
where the summation runs over all galaxies $j$ that belong to mass bin $i$. Accordingly, the counting error on the HIMF can be calculated as 

\begin{eqnarray}
\sigma_{\phi_i}^2 = \frac{1}{\Delta m_\mathrm{HI}^2} \cdot \sum_j \frac{1}{V_{\mathrm{eff},j}^2} \;\;\; .
\label{eq:veff_err}
\end{eqnarray}

\noindent
In the equations above, $\Delta m_\mathrm{HI}$ is the logarithmic width of the mass bin (i.e. $\Delta \log_{10}(M_\mathrm{HI}/M_\odot)$), while $V_{\mathrm{eff},j}$ is the `effective volume' available to galaxy $j$. The effective volume is determined through a maximum-likelihood statistical technique, and takes into account both the survey sensitivity limits and the fluctuations of galaxy counts with distance induced by the large-scale structure in the survey volume. As a result, the $1/V_\mathrm{eff}$ method is fairly robust against bias caused by inhomogeneities in the spatial distribution of galaxies. Full details of the implementation of the $1/V_\mathrm{eff}$ method in the context of the ALFALFA survey can be found in \citet[Appendix B]{Martin+2010} and \citet[\S 3.1]{Papastergis+2011}, and references therein.

There are two important technical differences between the measurement of the HIMF of various environmental subsamples in this work, and the measurement of the overall HIMF of ALFALFA \citep{Martin+2010}. First, it is very difficult to determine the actual survey volume occupied by each environmental subsample (see figure \ref{fig:SDSS_NN_cone}). As a result, we do not attempt to compute absolute normalisations for the environmental HIMFs, but rather we compare the HIMF shape among the various subsamples. Second, the spatial distribution of different environmental subsamples can be drastically dissimilar (see e.g. figure \ref{fig:FA_cone}). As a result, the effective volumes for galaxies that belong to a specific subsample are computed based on the spatial distribution of the other subsample members only (rather than the whole ALFALFA sample).

The method described above for the measurement of the HIMF is fully non-parametric. However, previous studies \citep[e.g.][]{Zwaan+2003,Zwaan+2005,Martin+2010} have shown that the HIMF can be described very well by a specific functional form, referred to as the `Schechter function' \citep{Schechter1976}:

\begin{eqnarray}
\phi(M_\mathrm{HI}) & = & \frac{dN_\mathrm{gal}}{dV \: d\log_{10}(M_\mathrm{HI})} = \nonumber \\
& = & \ln(10) \: \phi_\ast \: \left( \frac{M_\mathrm{HI}}{M_\ast} \right)^{\alpha+1} \: e^{-\left( \frac{M_\mathrm{HI}}{M_\ast}\right)} \;\;\; .
\end{eqnarray}

\noindent
The Schechter function describes a power law of logarithmic slope $\alpha+1$ at the low-mass end ($M_\mathrm{HI} \ll M_\ast$), which transitions to an exponential drop off at the high-mass end ($M_\mathrm{HI} \gg M_\ast$). The parameter $M_\ast$ is therefore the value of mass corresponding to the transition `knee' of the HIMF, while $\phi_\ast$ controls the normalisation of the HIMF. In this work, we determine the best fit Schechter parameters for the measured HIMFs by ordinary least squares minimisation\footnotemark{}. As explained in the previous paragraph, the value of $\phi_\ast$ in the environmental HIMFs is arbitrary, and only the two shape parameters ($M_\ast$ and $\alpha$) are physically relevant in this case. Note that the two shape parameters are covariant, such that the fit error is best depicted as an ellipse in the $\{M_\ast,\alpha\}$ plane. Lastly, keep in mind that the errors on the fit parameters depend on the errorbars of individual HIMF datapoints. These errorbars are computed through Eqn. \ref{eq:veff_err}, and represent the statistical counting error only. As a result, systematic uncertainties are not included in the fit error values quoted in this article. The robustness of the $1/V_\mathrm{eff}$ method is discussed further in appendix \ref{sec:HIMF_check}.

\footnotetext{The best fit parameters are determined by the \texttt{scipy.optimize.curve\_fit} routine written in the \texttt{Python} programming language. The minimisation is performed in linear space, assuming Gaussian errors with a magnitude determined by Eqn. \ref{eq:veff_err}.}

\section{Results}
\label{sec:results}

\subsection{SDSS Reference Catalogue}

The following subsection is concerned with the results obtained when defining an ALFALFA galaxy's environment based on the SDSS reference catalogue that extends from 500-15,500 \kms/$H_{0}$, this includes both the NN and FA methods for defining environment (see \S\ref{sec:env}).

\subsubsection{Nearest Neighbour Density}
\label{sec:NN3_results}

The nearest neighbour density calculated by the 3rd SDSS neighbour above the volume limiting absolute magnitude cut was used to define quartiles of environment for the ALFALFA galaxies. The galaxies from each quartile were used to calculate the HIMF for that environment (as described in \S\ref{sec:HIMFcalc}) and were compared to the HIMF calculated from all four quartiles combined.

\begin{figure*}
\centering
\includegraphics[width=\columnwidth]{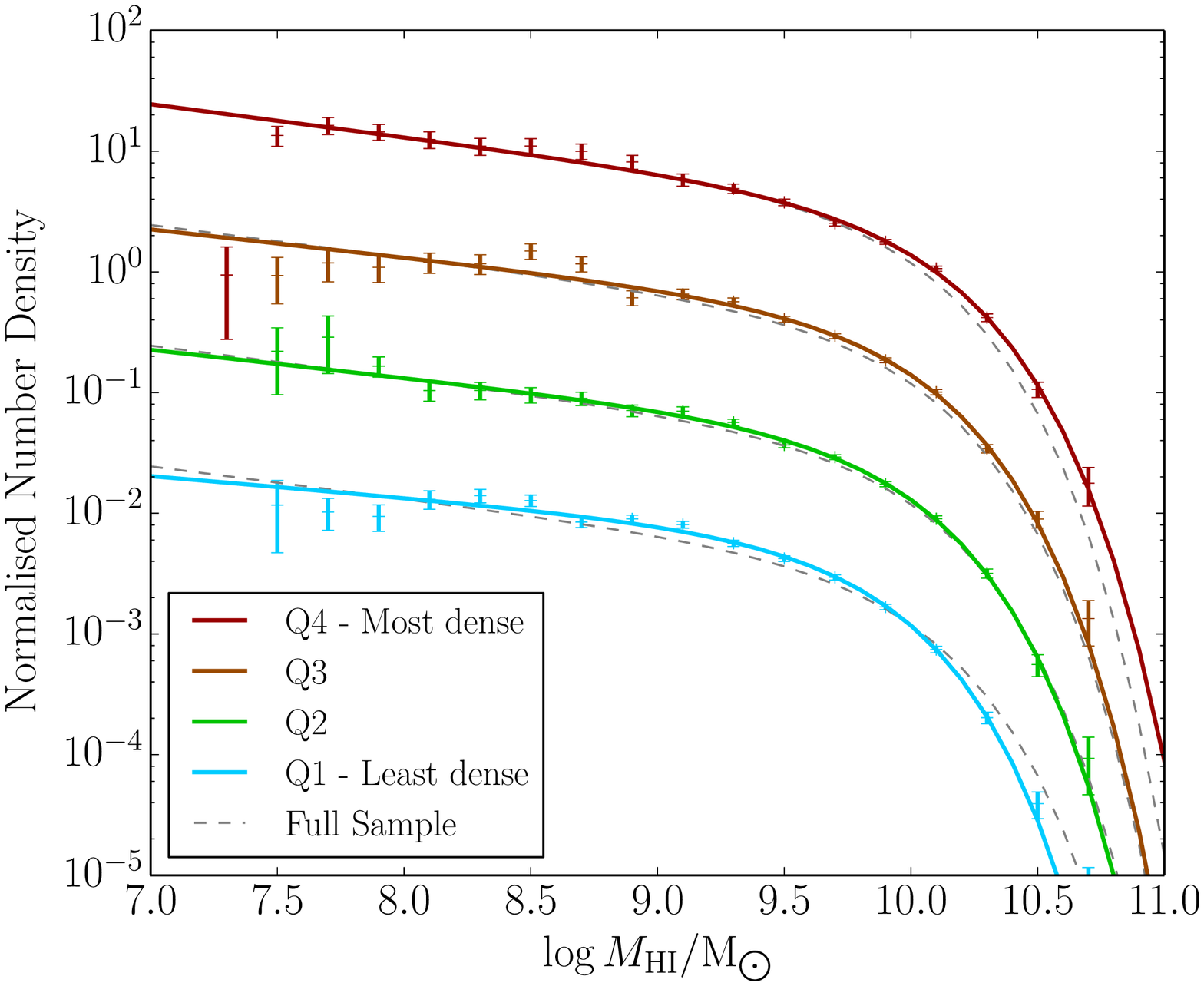}
\includegraphics[width=\columnwidth]{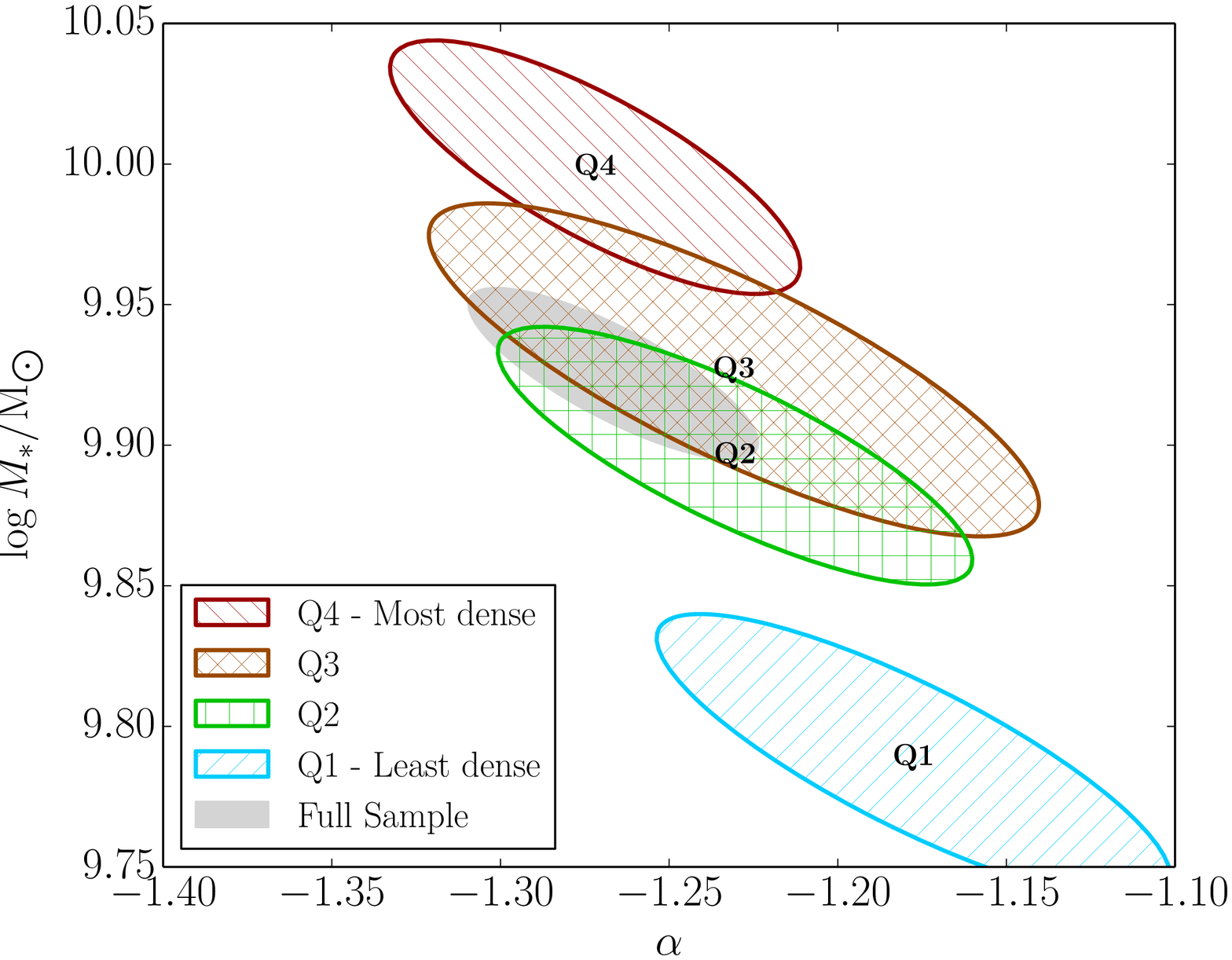}
\caption{\textit{Left panel}: The HIMFs of each environment density quartile in the ALFALFA sample. The solid coloured lines represent Schechter function fits of each quartile in nearest neighbour density, calculated using the 3rd nearest neighbour in the associated SDSS catalogue. In order of most to least dense they are dark red, gold, green, light blue, or equivalently, top to bottom (or dark to light shades). The dashed grey lines show the HIMF of the full sample, and are offset to aid readability. The error bars represent the counting errors only, and neglect errors in the input masses and velocity widths. \textit{Right panel}: The 2-$\sigma$ error ellipses of the Schechter function fit parameters of the HIMFs in the left plot. The colour scheme is identical to the left plot and the hatching styles are as follows: positively sloped, vertical cross, diagonal cross, negatively sloped, in order of increasing density quartiles. The grey filled ellipse represents the fit to the full sample.}
\label{fig:NN3_results}
\end{figure*}

Figure \ref{fig:NN3_results} shows the HIMF for each of the four ALFALFA quartiles (left) and the 2-$\sigma$ errors ellipses of the fit to the Schechter function parameters of each quartile (right). There is a clear trend of the lowest environmental density quartile (light blue) HIMF function falling below that of the full sample at the high mass end, and this switches to lying above it for the highest density quartile (dark red), with the middle two quartiles falling between the two extremes. There is also a much weaker dependence on the low-mass slope, with the quartiles appearing to produce a marginally flatter slope as the local density decreases.

Theses results seem to indicate that the `knee' mass of the HIMF is indeed a function of nearest neighbour environment (as defined by the SDSS reference catalogue in \S\ref{sec:nearest_neighbour}) with the value of $\log M_{*}/\mathrm{M_{\odot}}$ changing from $9.81 \pm 0.02$ to $10.00 \pm 0.03$ between the lowest and highest density quartiles (of the ALFALFA sample). There is also a suggestion of a trend in the low-mass slope, although this is much less pronounced. The error ellipses in figure \ref{fig:NN3_results} appear to move progressively further right (flatter low-mass slope) with decreasing density. However, this trend is not statistically significant as all the ellipses overlap in $\alpha$, indicating that they are consistent within 2-$\sigma$. Fitting a vertical line (fixed $\alpha$ value) to the ellipses results in a reduced $\chi^{2}$ value of 1.2, indicating that assuming no change in $\alpha$ is a reasonable model for the data (the equivalent $\chi^{2}$ value, assuming no change in $M_{*}$, is 13). It should also be noted that the Schechter fit is based only on the counting errors when calculating the HIMF, thus the error ellipses are likely underestimates of the true errors, as they do not include distance uncertainties (probably the largest single source of error). Furthermore, this apparent shift is in the direction that you would expect $\alpha$ to be driven by the change in $M_{*}$, due to the covariance between the two parameters. This is also opposite to the trend between environment and $\alpha$ that is expected (steeper in low density environments).

\subsubsection{Fixed Aperture Environment}
\label{sec:results_FA}

\begin{figure*}
\centering
\includegraphics[width=\columnwidth]{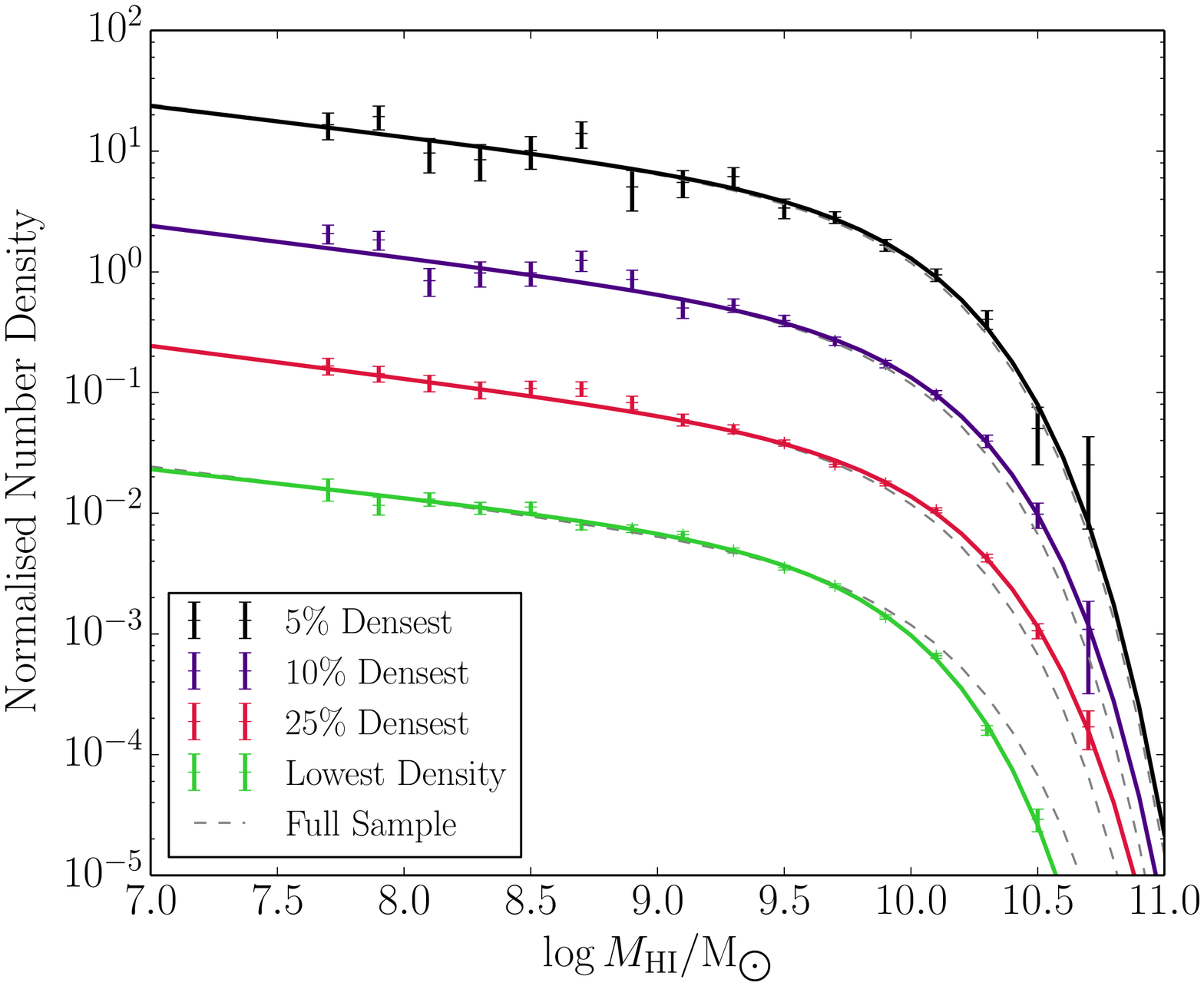}
\includegraphics[width=\columnwidth]{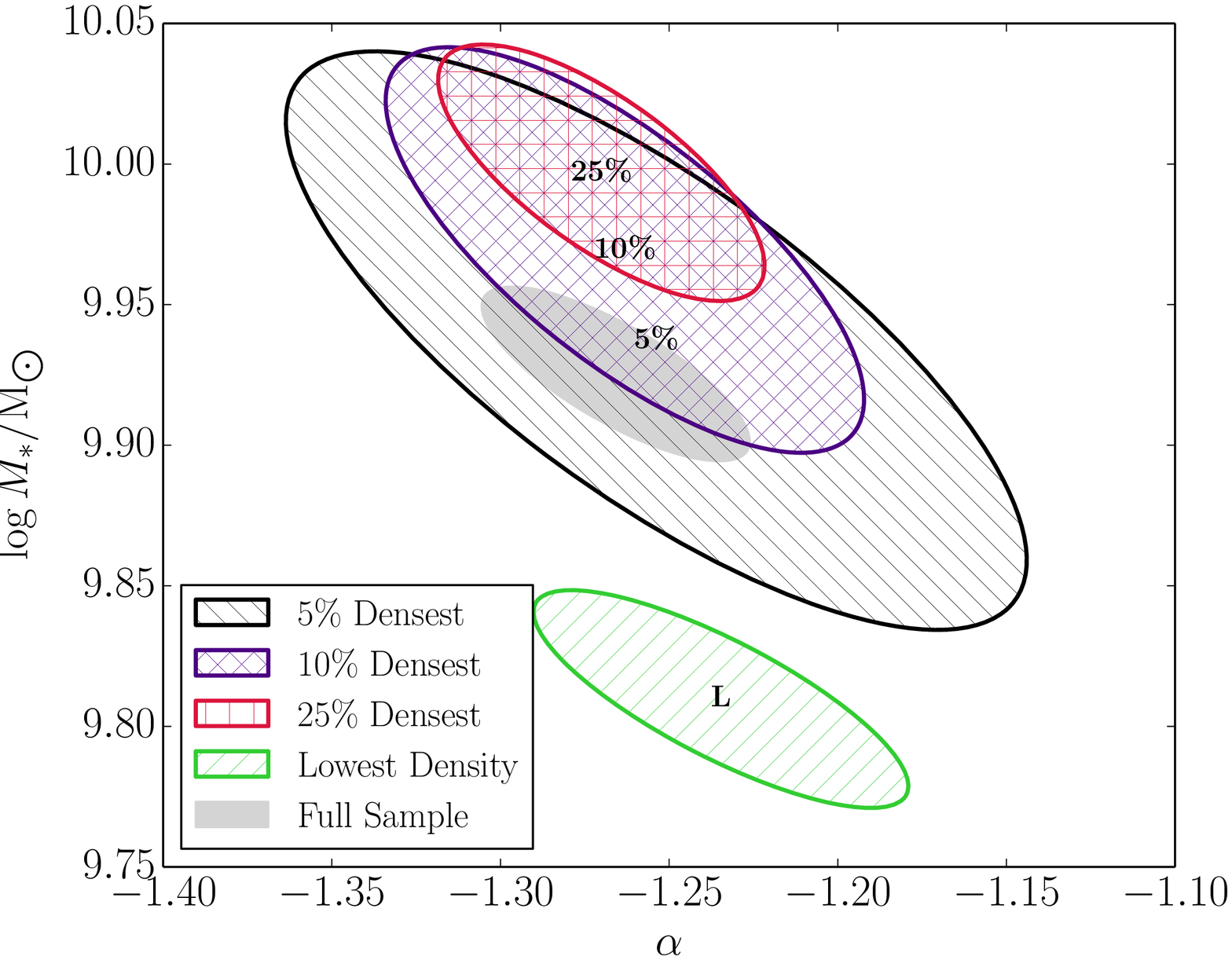}
\caption{\textit{Left panel}: The HIMFs of each environment defined by the fixed aperture method in the ALFALFA sample. The solid coloured lines represent the Schechter function fits of the four different environments, those with 0 neighbours within the fixed aperture, with 3 or more, 6 or more, or 9 or more. The respective colours are green, crimson, purple, and black or equivalently, bottom to top (or light shades to dark shades). The last three of these samples approximately corresponds to the 25, 10 and 5 percent most dense environments. The dashed grey lines show the HIMF of the full sample, and are offset to aid readability. The error bars represent the counting errors only. \textit{Right panel}: The 2-$\sigma$ error ellipses of the Schechter function fit parameters of the HIMFs in the left plot. The colour scheme is identical to the left plot and the hatching styles are as follows: positively sloped, vertical cross, diagonal cross, negatively sloped, in order of increasing density quartiles. The grey filled ellipse represents the fit to the full sample.}
\label{fig:HIMF_FA}
\end{figure*}

In Figure \ref{fig:HIMF_FA} we show the measured HIMFs and error ellipses for four environmental subsamples defined via the fixed aperture method (refer to \S\ref{sec:fixed_aperture}). In particular, the four sub-samples correspond to galaxies that belong to the lowest density FA environment (zero neighbours within the fixed aperture), and galaxies that belong to the 25\%, 10\% and 5\% densest environments in terms of FA neighbours. Figure \ref{fig:HIMF_FA} shows that there is no clear dependence of the low-mass slope on environment, in agreement with the findings of \S\ref{sec:NN3_results}. However, the environmental dependence of the `knee' mass is more complicated than before. In particular, we do observe a shift in the value of $M_\ast$ between the lowest density and 25\% densest FA sub-samples, that is compatible with the trend seen in figure \ref{fig:NN3_results}. However, the trend does not extend consistently to the two highest density FA sub-samples; instead the value of $M_\ast$ for the 10\% and 5\% densest FA sub-samples is actually slightly lower than for the 25\% sub-sample.

At first glance, the results of figures \ref{fig:NN3_results} and \ref{fig:HIMF_FA} regarding the environmental dependence of $M_\ast$ may seem inconsistent with each other. However, this is most probably not the case, because the two densest FA subsamples probe a higher density regime than the fourth quartile of NN environmental density (refer to \S\ref{sec:fixed_aperture}). We therefore interpret the observed $M_\ast$ trend with FA environment as the result of HI-deficiency affecting galaxies in the highest density regions of the ALFALFA volume. According to this interpretation, the extrapolation of the environmental $M_\ast$ trend observed for the NN subsamples into the highest density environments fails, because the processes responsible for HI-deficiency inhibit the formation of galaxies with high HI masses in these crowded environments.

\subsection{2MRS Reference Catalogue}
\label{sec:2MRS_NN}

\begin{figure*}
\centering
\includegraphics[width=\columnwidth]{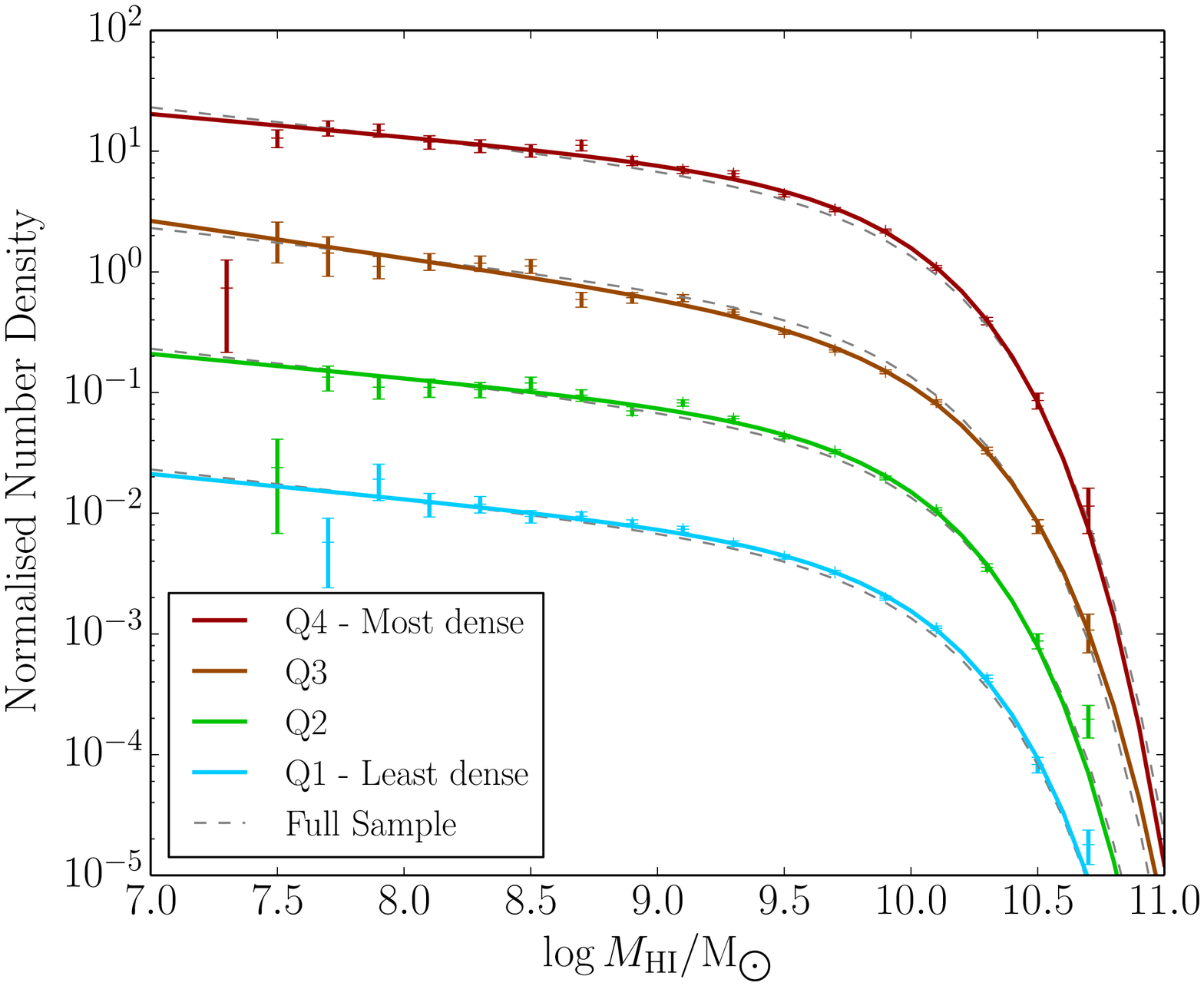}
\includegraphics[width=\columnwidth]{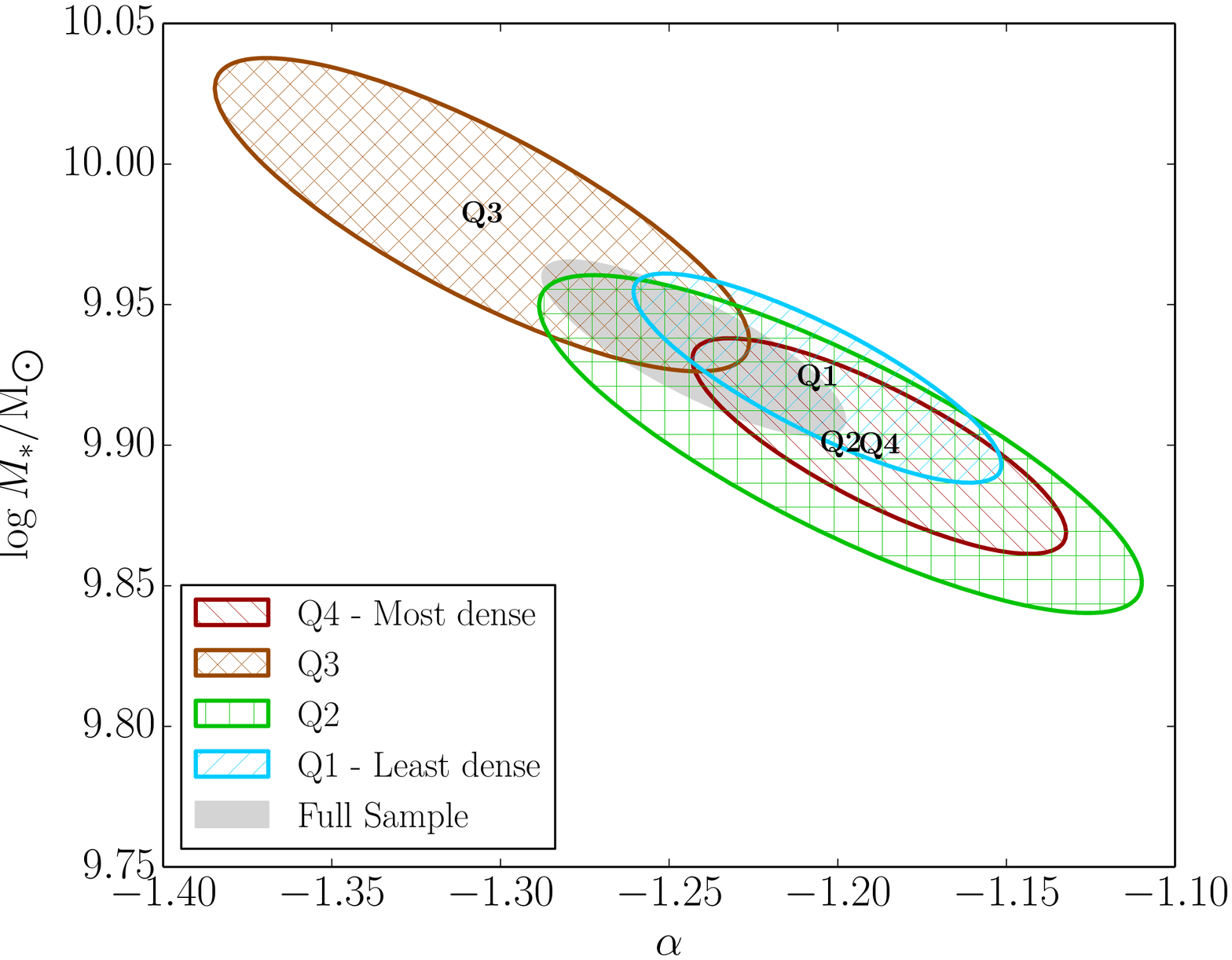}
\caption{Identical to figure \ref{fig:NN3_results} except that here nearest neighbour environment quartiles are defined using the first neighbour in the 2MRS reference catalogue.}
\label{fig:HIMF_2MRS_NN}
\end{figure*}

The 1st nearest neighbour in the volume-limited 2MRS catalogue was used to define quartiles of environmental density for the ALFALFA galaxies (refer to \S\ref{sec:2MRS}). 
The ALFALFA sample that can be used in the 2MRS analysis contains about 50\% more galaxies than the sample used in the SDSS analysis, as 2MRS is all sky survey. Figure \ref{fig:HIMF_2MRS_NN} shows the HIMF Schechter parameters calculated for each quartile of neighbour density (in 2MRS). Despite having a greater number of sources to compute the HIMFs, and therefore smaller error ellipses, no consistent trend in either $M_{*}$ or $\alpha$ is evident; all four quartiles are consistent with the global sample at 2-$\sigma$ confidence. This result has been checked to be robust against cosmic variance and the colour of the reference sample (see appendix for details).

The fundamental difference between the SDSS-based and 2MRS-based environmental measures is the scales that they probe. As argued in \S\ref{sec:2MRS}, the environment defined using 2MRS is probing a larger scale than that defined using SDSS. This is because 2MRS is a shallower survey, which leads to larger separations between sources. In addition, using the 2MRS catalogue to define the environment results in a better separation of our ALFALFA sample based on the position of galaxies in the LSS. For example, filaments and clusters are starkly defined in figure \ref{fig:2MRS_NN_cone} (right panel), while in the left panel there are clear gaps in the corresponding positions. Given these differences between environment defined using SDSS and 2MRS, and the fact that a trend between environment and $M_{*}$ is only measured when using SDSS, the most straightforward interpretation of our results is that an HI-selected galaxy's characteristic HI mass ($M_{*}$) increases with the density of its local environment, but is independent of its position relative to large scales structures. In addition, we find that the faint end slope of HI-selected galaxies is universal, having no significant dependence on any measure of environment we explored.

\section{Discussion}
\label{sec:discuss}

The notion that local environment is the primary factor for determining a galaxy's properties is not a new idea, in fact it is the fundamental assumption underlying the very successful HOD formalism. There are also optical based experiments which have found similar results: \citet{Berlind+2005} compared simulations and the SDSS to demonstrate that galaxy properties are strongly correlated with the host halo mass, and that this is the parameter that most environment measures based on local galaxy density, are tracing; \citet{Blanton+2006} studied the environment of SDSS galaxies on different scales and found that only environment within $\sim$1 Mpc is important for determining a galaxy's star formation rate and colour. Our results fit well with these theoretical and optical results, however there are still a number of tensions with theory and other HI observations. Below we review some literature results regarding the environmental dependence of the HIMF and discuss cases where there exists tension between these studies and the results of this work.

\subsection{Comparison with previous HI survey results}
\label{sec:discussion_surveys}

The first study on the environmental dependence of the HIMF based on a large-area blind HI survey was performed by \citet{Zwaan+2005}, using the HIPASS dataset. Contrary to our results, they found that the low-mass slope, $\alpha$, becomes steeper with increasing environmental density, while the `knee' mass, $M_\ast$, is roughly independent of the environment (see their figure 3). The comparison between the HIPASS result of \citet{Zwaan+2005} and the ALFALFA result obtained in \S\ref{sec:results} is not straightforward, because the two studies define the NN environment in different ways. In particular, \citet{Zwaan+2005} find neighbours for the HIPASS galaxies in the HIPASS catalogue itself. This decision was dictated by the fact that there is no large-area spectroscopic survey at optical wavelengths that covers the HIPASS footprint. As explained in \S\ref{sec:reference_catalogue}, this neighbour definition makes the consistent computation of environmental density throughout the survey volume very difficult to achieve in practice. In Appendix \ref{sec:HIPASS_mock} we show that if an environmental trend in $\alpha$, equivalent to that found in HIPASS by \citet{Zwaan+2005}, was present in the ALFALFA dataset, it would have been easily detected by the current analysis.  

Another important difference between the HIPASS and ALFALFA nearest neighbour definitions is the scale over which they probe the environment. More specifically, the HIPASS catalogue is much sparser than the SDSS reference catalogue used for environment definition in this work. If not due to the observational limitations therefore, the HIPASS trend should be driven by the large-scale environment of galaxies, rather than the local environment probed in this article. However, this interpretation of the HIPASS result is also open to question. For example, \citet{Moorman+2014} have recently measured the HIMF separately for the ALFALFA galaxies that reside in voids and for those that reside in walls/filaments. They find a difference in the HIMF measured for the two environmental samples that is similar to the environmental trend found in \S\ref{sec:NN3_results}. In particular, the wall/filament HIMF has a higher `knee' mass than the void HIMF, but only a marginally steeper low-mass slope (refer to their figure 8). Given that the \citet{Moorman+2014} environment definition also refers to large scales ($\sim$10 Mpc), their result seems to contradict the HIPASS finding.

Our results make an intriguing addition to those of \citet{Moorman+2014} because we detect a very similar trend in $M_{*}$, but associated with local, rather than large scale, environment. The reason for this apparent contradiction is not clear, however we note that it could be resolved if the separation of galaxies between void and wall objects in \citet{Moorman+2014} is correlated with the local environment of the galaxies more than naively expected based on the size of these cosmic structures; in that case, it would be natural for the \citet{Moorman+2014} result to be closely related to the result obtained by considering SDSS-based local densities.

An additional complication is added by the fact that the \citet{Moorman+2014} trend is not detected in the present work when environment is defined on relatively large scales with 2MRS-based densities (see \S\ref{sec:2MRS_NN}). Again, the reason for this tension is not entirely clear, although (as above) if the void and wall samples of \citet{Moorman+2014} were sufficiently correlated with local density, then a trend associated with local environment could be masquerading as one with large scale environment -- a false trend that we would not necessarily expect to see with 2MRS neighbour densities. Alternatively, it is possible that 2MRS could be missing the large scale component of a real trend associated with both local and large scale. 2MRS clearly separates out the densest LSS into the 4th quartile of neighbour density, but if the separation between the remaining 3 quartiles was extremely noisy, then trends could be suppressed.

\subsection{Comparison to the HIMF in groups}
\label{sec:group_HIMF}

\citet{Verheijen+2001, Kovac+2005, Freeland+2009, Pisano+2011} studied the HIMF in galaxy groups and all came to essentially the same conclusion; that the low-mass slop is flat in groups. Given these consistent findings, it is perplexing that we see no evidence for variation of the low-mass slope, as in the field it has been shown by both HIPASS \citep{Zwaan+2005} and ALFALFA \citep{Martin+2010} that it is not flat (both surveys measure $\alpha \approx -1.3$). 

Assuming that a non-negligible fraction of ALFALFA's detections are galaxies in groups \citep[][find that approximately 25\% of ALFALFA galaxies are in groups]{Hess+2013}, such that any trend would not be drowned out, then the findings above suggest that the nearest neighbour definition of environment is not consistently separating groups from the rest of the sample. If this were not the case, then there would need to be an inconsistency in how the wide field and targeted surveys are calculating the HIMF, in order to explain these seemingly contradictory findings. 

This apparent shortcoming in the nearest neighbour method could be explained if the surface number density of galaxies in groups is approximately independent of group size. As our method cannot distinguish regions of the same surface density, under these assumptions, it would be incapable of separating groups of different sizes and we would be blind to any trend associated with group size. Therefore, if the low mass slope varies with group size, our analysis might not reveal this. Alternatively, as the surveys which have measured a flat low-mass slope in groups are mostly interferometric surveys (that resolve many of their sources), an uncertain detection threshold associated with HI surface density could result in an erroneous slope. A more detailed study of the HIMF in groups is required to test these hypotheses and compare the two existing methodologies.

\section{Conclusions}
\label{sec:conclude}

We have used the 70\% ALFALFA sample to search for dependence of the HIMF on galactic environment. In particular, we defined the environment of ALFALFA galaxies based on the neighbours found in both SDSS and 2MRS volume limited reference catalogues. We find that the Schechter function `knee' mass ($\log{M_{*}/\mathrm{M_{\odot}}}$) is dependent on environment, with its value shifting from $9.81 \pm 0.02$ to $10.00 \pm 0.03$ between the lowest and highest density quartiles. However, this dependence was only observed when defining environment based on the SDSS reference catalogue, not 2MRS. Using a fixed aperture measure of environment with SDSS, we also found tentative evidence for a decrease in $M_{*}$ in the highest density environments, in agreement with the notion that galaxies in clusters should become HI-deficient.

In \S\ref{sec:env} we demonstrated that using our approach, 2MRS both measures environment on a larger scale than SDSS, and is more effective at separating large scale structures into different environment density quartiles. This strongly suggests that the dependence we are seeing is on local environment, rather than large scale, supporting the fundamental assumption of the HOD formalism, that a galaxy's properties are only dependent on the mass of its host halo. However, this is in tension with a previous ALFALFA-based study \citep{Moorman+2014} which found a similar trend in $M_{*}$, but based on separating galaxies which reside in walls and voids. 

Although the true resolution remains unclear we offered two potential explanations for this discrepancy between our results and those of \citet{Moorman+2014}. If void and wall environments are sufficiently correlated with local densities such that trends are expected with either definition of environment, then the results would be in agreement. Alternatively, if the 2MRS densities used in this paper were to be incapable of distinguishing low density environments then trends associated with large scales might be hidden from our analysis.

In all of the tests we performed we detected no significant dependence of the the low-mass slope ($\alpha$) on environment. Again, this appears in conflict with existing results, both from HIPASS \citep{Zwaan+2005} and from several studies of galaxy groups (which measure $\alpha \sim -1$). The steepening of $\alpha$ with denser environments that was observed in HIPASS is not directly comparable to this article due to different methodology (see \S\ref{sec:reference_catalogue}), and in appendix \ref{sec:HIPASS_mock} we demonstrate that we would be capable of detecting an equivalent trend if it existed in our data. As an explanation to resolve the tension with the findings of group HI studies, we suggest that the inability of the nearest neighbour environment to separate different sized groups of the same projected surface density, might be responsible for our null result. If the low-mass slope was a function of group size and most groups had similar surface densities, then this would explain the observations. An alternative explanation could be inconsistent methodologies resulting from uncertain surface brightness limits in narrow field surveys. A more complete understanding of the HIMF in groups is needed to test these hypotheses.

\section*{Acknowledgements}

The authors acknowledge the work of the entire ALFALFA collaboration in observing, flagging, and extracting the catalogue of galaxies that this paper makes use of. The ALFALFA team at Cornell is supported by NSF grants AST-0607007 and AST-1107390 to RG and MPH and by grants from the Brinson Foundation. EP is supported by a NOVA postdoctoral fellowship at the Kapteyn Institute. MGJ would like to thank Kelley Hess for useful discussions about HI in groups.

\bibliography{refs}

\appendix
\section{Robustness of Results}
\subsection{Impact of Confusion}

In \citet{Jones+2015} we warned that source confusion acts to increase the observed value of $M_{*}$, and that as confusion is undoubtedly a function of environment, caution must be used when looking for environmental trends in that parameter. The shift observed here is approximately 0.2 dex, whereas in \citet{Jones+2015} the maximum shift created by confusion in an ALFALFA-like survey was estimated to be 0.06 dex. This indicates that confusion is very unlikely to be the source of the trend detected.

\subsection{Testing Cosmic Variance}

To test whether the results were biased by the particular directions the ALFALFA survey mapped we carried out two additional tests. The first was to simply eliminate the Virgo cluster. The Virgo cluster is the most dominant nearby, large scale feature in the Spring sky (where there is also SDSS coverage), and due to its proximity it raises the question of whether we can detect systems there that we would never see if it were at a distance more typical of the other clusters in the survey, and whether this might bias our results. By removing all sources with recession velocities less than 3,000 \kms, we remove almost all those objects associated with Virgo, and repeat our analysis.

Though clipping the inner 3,000 \kms \ of the data severely impacts our ability to constrain the low-mass slope, the trend in the `knee' mass is still preserved, indicating that the exceptional location of Virgo is not the driving force behind this result.

The second test of cosmic variance involves comparing the Spring and Fall skies. As there is little spectroscopic SDSS coverage in Arecibo's Fall sky, this must rely on comparison with 2MRS. As shown in \S\ref{sec:2MRS_NN} when all of ALFALFA 70\% is considered with environment defined using 2MRS there is no apparent trend in $M_{*}$ or $\alpha$ with environment. This could either be due to the previous trend being a property only of the Spring sky (cosmic variance) or due to the differences between 2MRS and SDSS as reference catalogues. To assess which it was the same analysis was repeated again, but now only considering ALFALFA sources in the Spring sky. As before, when environment is defined by 2MRS, there is no apparent trend in $M_{*}$ or $\alpha$, suggesting that the trend observed with SDSS is indeed real and not due to cosmic variance, and that that apparent lack of such a trend with environment define by 2MRS is due to differences in the galaxies detected in those two surveys.

\subsection{Independence and Covariance}

In order to ensure that our findings are not dependent on the exact magnitude limits we set to make the SDSS and 2MRS catalogues volume limited, we repeated our analysis with three additional samples with ranges of 1,000-6,000, 1,000-8,000, and 1,000-12,000 \kms $/H_{0}$ in ALFALFA, and an additional 500 \kms $/H_{0}$ at either edge in SDSS and 2MRS. The trend in $M_{*}$ shown in figure \ref{fig:NN3_results} appears in all three additional samples when compared to SDSS (but not any 2MRS reference sample), although the error ellipses become progressively larger as the samples get smaller. The apparent shift in $\alpha$ is also persistent across all the different ALFALFA samples, suggesting that it is statistically significant. However, not only should caution be used because $\alpha$ and $M_{*}$ are highly covariant (and the assumed Gaussian errors likely do not fully encompass this dependence), but the four samples themselves are not independent because the galaxies in the 1,000-6,000 \kms \ catalogue are contained within the other three. While this latter point is unimportant for fitting $M_{*}$, as those galaxies are detectable by ALFALFA throughout all the samples, the galaxies in the 1,000-6,000 \kms \ catalogue dominate the low-mass population. Therefore if the trend in $\alpha$ with environment is not significant in the main sample (as was shown in \S\ref{sec:discuss}), then it is not significant.

We also wished to check that the difference between the SDSS and 2MRS definitions of environment were due to the scale probed and not the colour of the reference population. To do this the SDSS catalogue was split into blue and red sub-catalogues, with the division occurring at $u-r = 2.2$. The third nearest neighbour environment was then re-calculated for the red and blue reference catalogues separately. The trend in $M_{*}$ remained significant in both samples, though the range of $M_{*}$ was marginally larger when environment was defined by the blue population. The modal values of the neighbour densities were equivalent between the red and blue nearest neighbour environments, with the distribution of densities defined by the red population showing a slightly longer tail towards high density.

\section{HIPASS Low-mass Slope Trend}
\label{sec:HIPASS_mock}

Other than studies based on ALFALFA, the largest sample used to study the variation of the HIMF with environment was carried out with the HIPASS dataset \citep{Barnes+2001}. In particular, \citet{Zwaan+2005} found that the HIMF low-mass slope, $\alpha$, becomes steeper in higher density environments (see their figure 3). On the other hand, they found no trend of $M_\ast$ with environmental density. In this section, we try to assess whether the presence of such an environmental trend would be detectable in the 70\% ALFALFA sample. In order to do that, we first divide the sample into five equally spaced logarithmic bins in local density. We quantify the local density as described in \S\ref{sec:nearest_neighbour}, i.e. by considering the 3$^\mathrm{rd}$ nearest neighbour in the SDSS reference catalogue. Note that the environmental division scheme considered here tries to reproduce the one used in \citet{Zwaan+2005}, and it is not the same as the one used to create our Figure \ref{fig:NN3_hist}. In particular, here we split galaxies into five logarithmic density bins of equal width, while Figure \ref{fig:NN3_hist} refers to four environmental sub-samples defined such that each contains the same number of objects.

We then create five mock samples, each consisting of approximately the same number of objects as one of the five real environmental subsamples in ALFALFA. The five mock samples are further created to reproduce the large-scale structure observed for their corresponding ALFALFA subsample. The mock samples mimic the environmental dependence of the HIMF found by HIPASS: the low-mass slope ranges from $\alpha = -1.2$ for the lowest density subsample to $\alpha = -1.52$ for the highest density one, while the `knee' mass is kept fixed at $\log(M_\ast/M_\odot) = 9.94$.

\begin{figure*}
\centering
\includegraphics[width=\columnwidth]{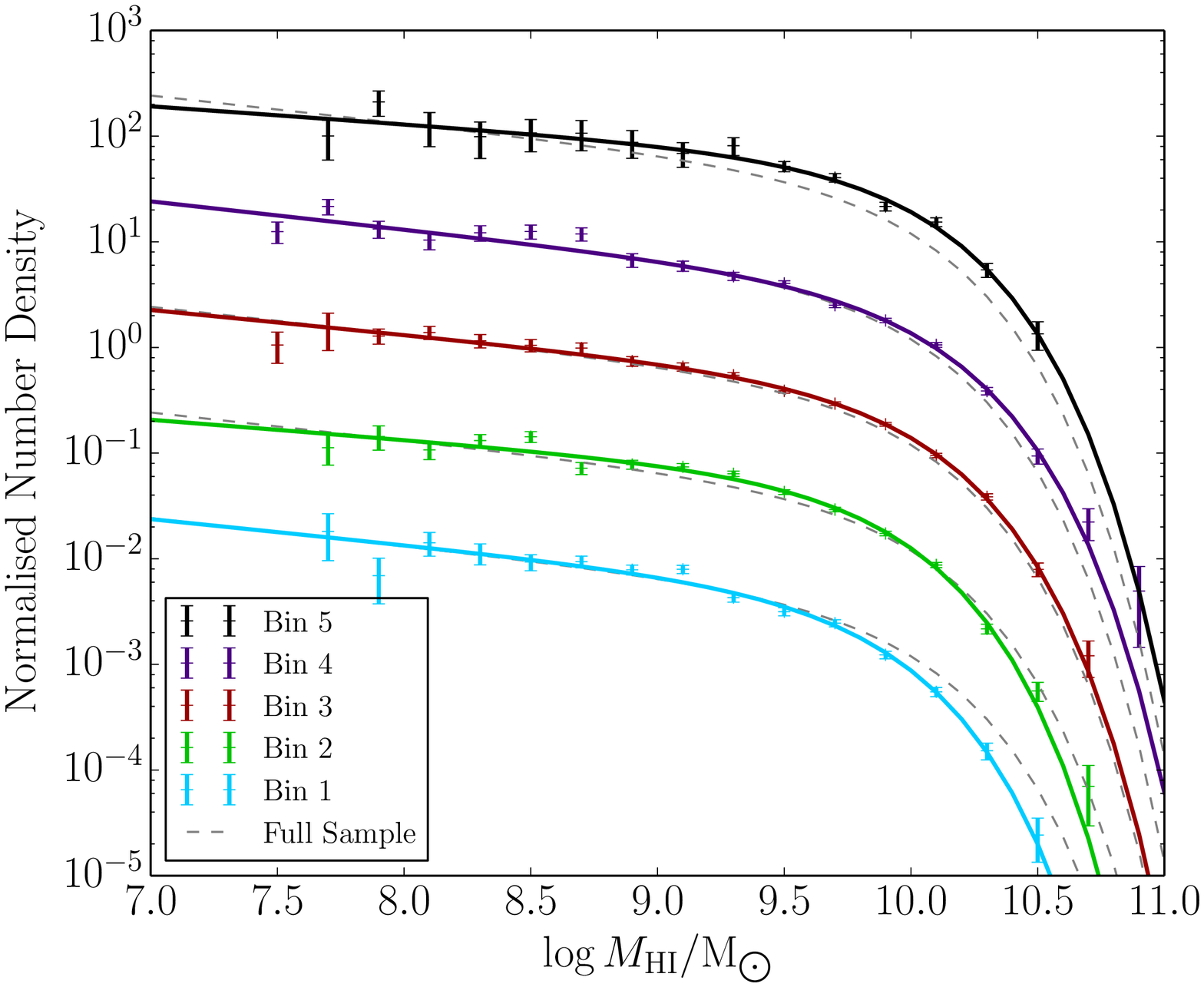}
\includegraphics[width=\columnwidth]{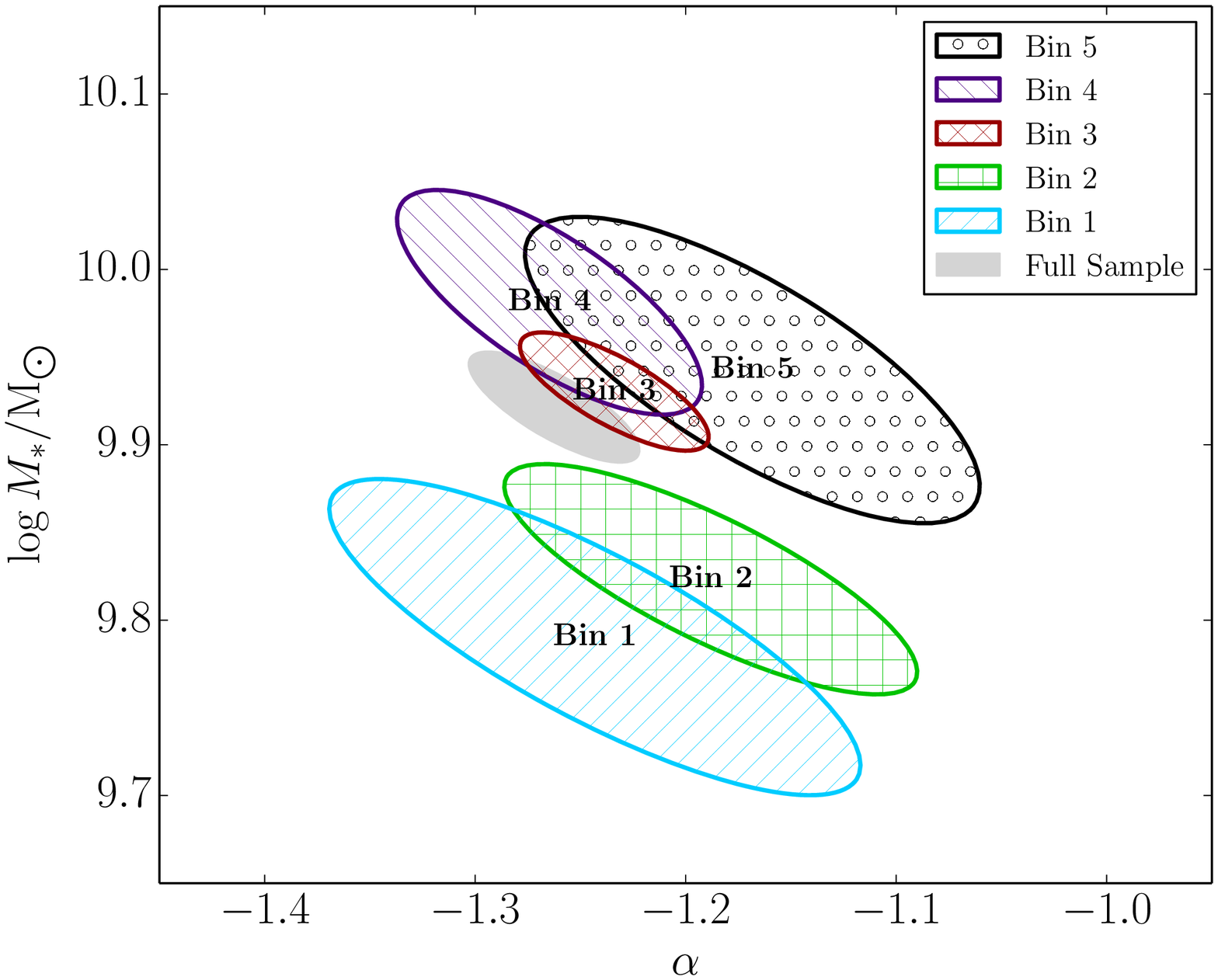}
\includegraphics[width=\columnwidth]{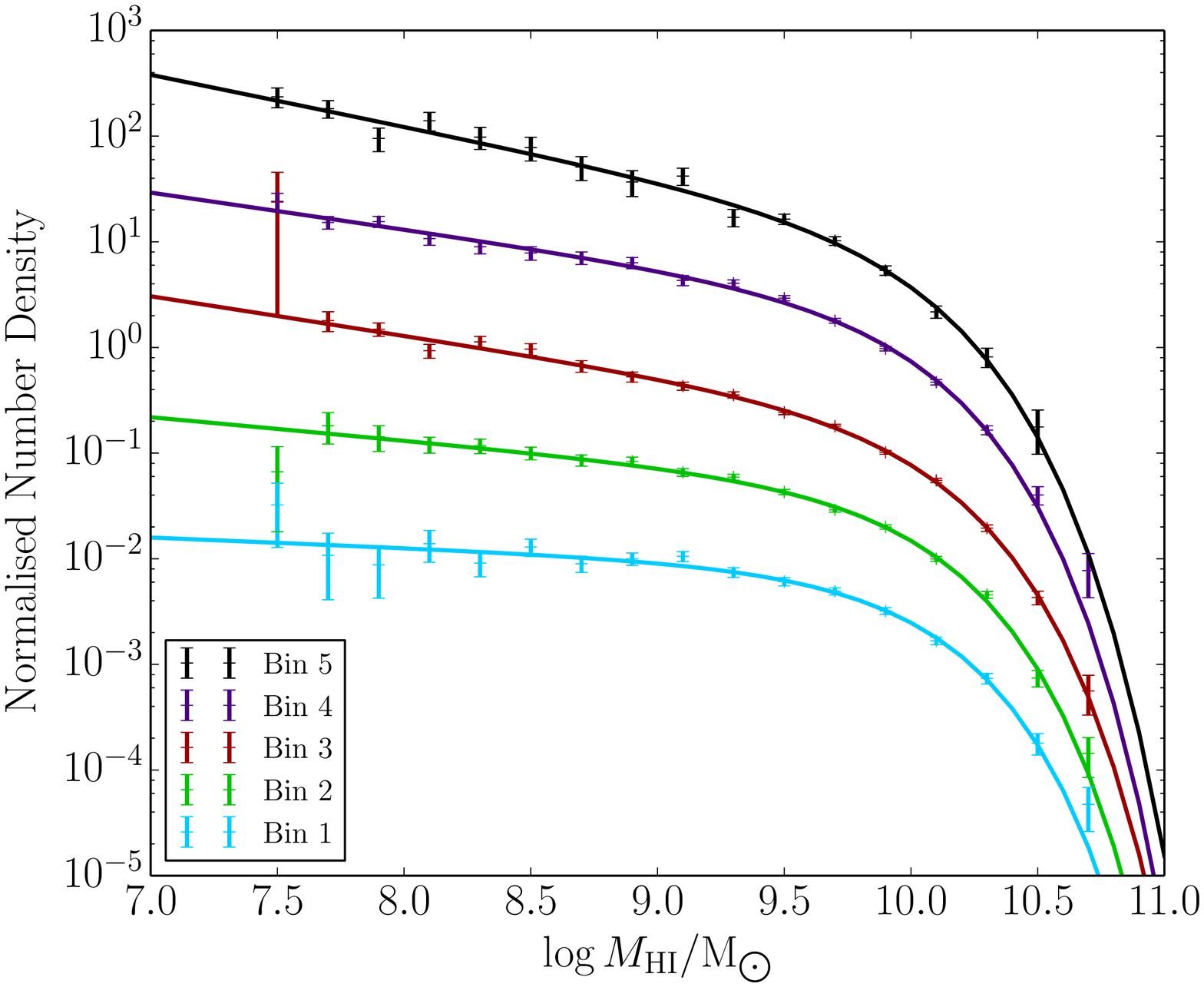}
\includegraphics[width=\columnwidth]{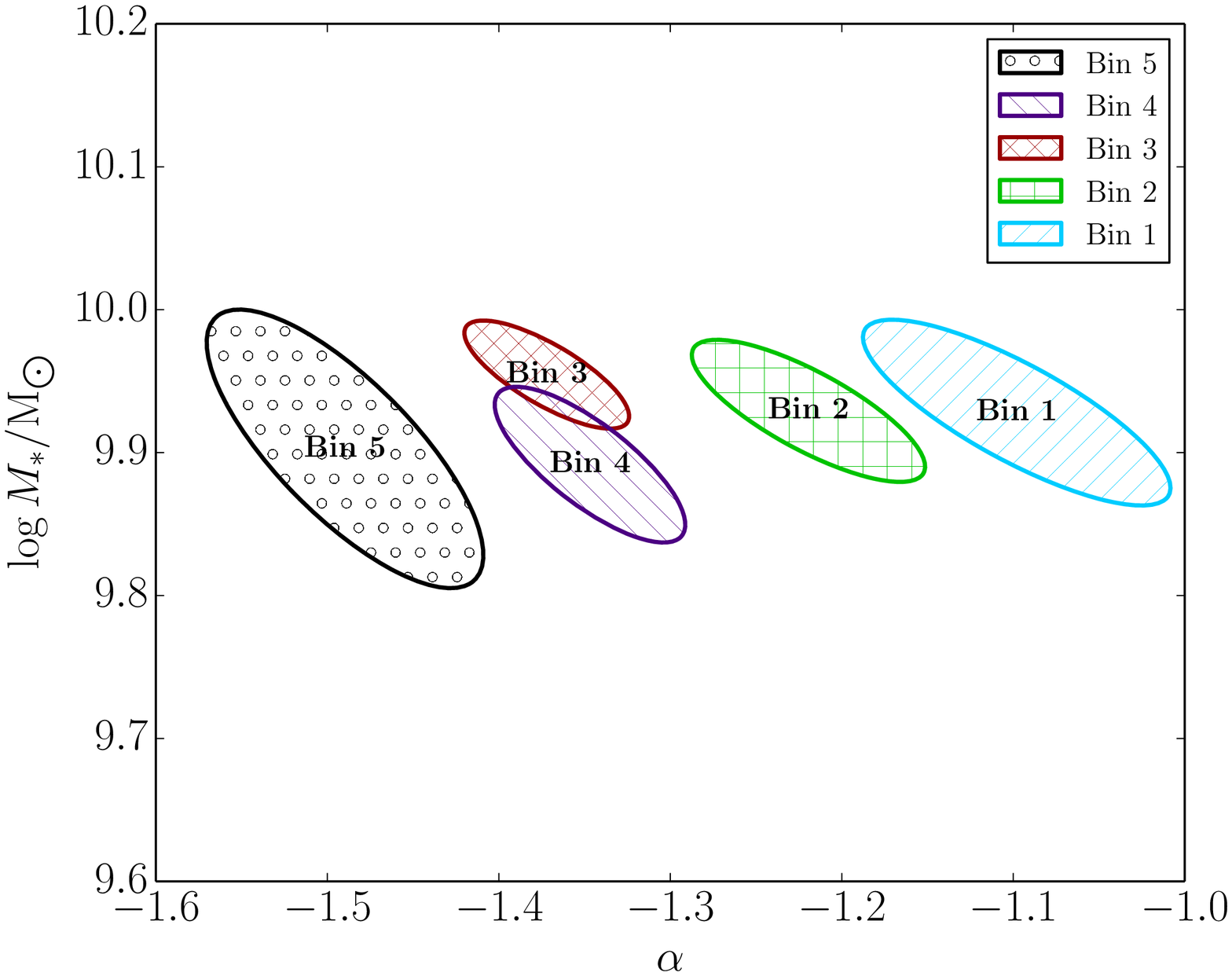}
\caption{
\textit{Top row}: HIMFs (\textit{left panel}) and Schechter parameter error ellipses (\textit{right panel}) for the five environmental ALFALFA subsamples defined in Appendix \ref{sec:HIPASS_mock}. The layout of the figure follows the layout of figure \ref{fig:NN3_results}, where HIMF colours from black to light blue (dark shades to light shades, and HIMF positions from top to bottom) correspond to subsamples of progressively lower environmental density. \textit{Bottom row}: Same as the top row, but for the five mock environmental subsamples, which are created to mimic the HIMF trends reported by \citet{Zwaan+2005} based on the HIPASS dataset (see Appendix \ref{sec:HIPASS_mock} for details).
The figure demonstrates that, if the HIMF trend claimed by \citet{Zwaan+2005} were present in the ALFALFA sample, it would have been easily detected.
}
\label{fig:HIPASS_mock}
\end{figure*}

We measure the HIMF in each of the five real and each of the five mock environmental subsamples, always using the same methodology (refer to Sec. \ref{sec:HIMFcalc}). The result is shown in Figure \ref{fig:HIPASS_mock}: The upper row shows the five HIMFs for the real environmental subsamples in ALFALFA, while the bottom row shows the HIMFs measured for the five mock samples. As expected based on the results of \S\ref{sec:NN3_results}, we see in the top row a clear trend of increasing $M_\ast$ value in higher density environments, and no significant environmental trend in $\alpha$. On the other hand, the bottom row reproduces very well the HIPASS-like environmental trend used to create the five mocks; the low-mass slope becomes steeper with increasing environmental density, while there is no significant environmental dependence of $M_\ast$. Figure \ref{fig:HIPASS_mock} demonstrates that if an environmental dependence of the HIMF similar to what measured by HIPASS were indeed present in the 70\% ALFALFA catalogue, it would have easily been detected in our current analysis.

At the same time, keep in mind that the comparison between the HIPASS and ALFALFA results is subject to one caveat. In particular, the HIPASS trend was detected when the environment was defined in terms of the 3$^\mathrm{rd}$, 5$^\mathrm{th}$ and 10$^\mathrm{th}$ nearest HIPASS neighbour \citep[Figure 4 in][]{Zwaan+2005}. However, the HIPASS HI-selected sample is much sparser than the SDSS reference catalogue used to calculate 3$^\mathrm{rd}$ nearest neighbour densities for ALFALFA galaxies. It cannot be excluded therefore that the trend observed by HIPASS is present only when environment is defined on very large scales, but is absent when local environment is considered. Please refer to \S\ref{sec:discussion_surveys} for a more thorough discussion of the tension with the HIPASS result.

\section{Robustness of the HIMF to LSS}
\label{sec:HIMF_check}

There is substantial deviation from a uniform distribution of galaxy positions in the ALFALFA sample due to the LSS in the nearby Universe, therefore it is important to confirm that the $1/V_\mathrm{eff}$ method for calculating the HIMF (see \S\ref{sec:HIMFcalc}) is robust to such deviations. To check this we used the methodology of \citet{Jones+2015} to create a uniform mock catalogue of approximately 15,000 HI sources. A second catalogue was produced by adding Gaussian overdensities of sources at distances of 20 and 100 Mpc, in order to simulate the Virgo cluster at the Great Wall, two major structures in the ALFALFA footprint. Finally, a third catalogue was made with sources removed from around 100 Mpc to create an effective void. Each of these three mocks was generated 30 times with input Schechter function shape parameters $\alpha = -1.30$ and $M_{*} = 9.95$. The mean values derived from the samples using the $1/V_\mathrm{eff}$ method were $\alpha = -1.28 \pm 0.01$ and $M_{*} = 9.96 \pm 0.01$, $\alpha = -1.27 \pm 0.01$ and $M_{*} = 9.96 \pm 0.01$, $\alpha = -1.27 \pm 0.01$ and $M_{*} = 9.96 \pm 0.01$, for the uniform mock, mock with overdensities, and the mock with a void, respectively. Similar experiments were carried out for a variety of input parameters ($\alpha$ and $M_{*}$) and gave equivalent results.

These results illustrate two important points: a) the $1/V_\mathrm{eff}$ method appears to be very robust against density deviations along the light of sight, b) the method has a slight systematic bias towards flatter low-mass slopes at the level of the second decimal place. This bias in the maximum likelihood methods has been known for some time \citep[e.g.][]{Efstathiou+1988,Willmer1997} and is not presently a major concern given our level of precision, however for future surveys with larger datasets a different estimator may be required.

\end{document}